\documentclass{svmult}

\usepackage{mathptmx}       
\usepackage{helvet}         
\usepackage{courier}        
\usepackage{type1cm}        
\usepackage{makeidx}         
\usepackage{graphicx}        
\usepackage{multicol}        
\usepackage[bottom]{footmisc}

\usepackage{subcaption}
\usepackage{array}
\newcolumntype{M}[1]{>{\centering\arraybackslash}m{#1}}

\usepackage{booktabs}
\usepackage{amsfonts}
\usepackage{siunitx}
\usepackage{soul}
\usepackage{wrapfig}
\usepackage{enumitem}

\makeindex             

\begin{document}

\title*{On the Social Influence in Human Behavior: Physical, Homophily, and Social Communities}
\author{Luca Luceri, Alberto Vancheri, Torsten Braun, Silvia Giordano}
\institute{Luca Luceri, Alberto Vancheri, and Silvia Giordano \at University of Applied Sciences and Arts of Southern Switzerland (SUPSI), Via Cantonale 2C, 6928 Manno, Switzerland, \email{name.surname@supsi.ch}
\and Luca Luceri, and Torsten Braun \at University of Bern, Neubrückstrasse 10, 3012 Bern, Switzerland, \email{luca.luceri@students.unibe.ch, braun@inf.unibe.ch}
\and This work is supported by the Swiss National Science Foundation via the SwissSenseSynergy project, grant number 154458.
}
%

\titlerunning{On the Social Influence in Human Behavior}
\maketitle

\abstract*{Understanding the forces governing human behavior and social dynamics is a challenging problem.
Individuals' decisions and actions are affected by interlaced factors, such as physical location, homophily, and social ties.
In this paper, we propose to examine the role that distinct communities, linked to these factors, play as sources of social influence. 
The ego network is typically used in the social influence analysis.
Our hypothesis is that individuals are embedded in communities not only related to their direct social relationships, but that involve different and complex forces.
We analyze physical, homophily, and social communities to evaluate their relation with subjects' behavior.
We prove that social influence is correlated with these communities, and each one of them is (differently) significant for individuals. We define community-based features, which reflect the subject involvement in these groups, and we use them with a supervised learning algorithm to predict subject participation in social events.
Results indicate that both communities and ego network are relevant sources of social influence, confirming that the ego network alone is not sufficient to explain this phenomenon.
Moreover, we classify users according to the degree of social influence they experienced with respect to their groups, recognizing classes of behavioral phenotypes.
To our knowledge, this is the first work that proves the existence of phenotypes related to the social influence phenomenon.
}

\abstract{Understanding the forces governing human behavior and social dynamics is a challenging problem. 
Individuals' decisions and actions are affected by interlaced factors, such as physical location, homophily, and social ties.
In this paper, we propose to examine the role that distinct communities, linked to these factors, play as sources of social influence. 
The ego network is typically used in the social influence analysis.
Our hypothesis is that individuals are embedded in communities not only related to their direct social relationships, but that involve different and complex forces.
We analyze physical, homophily, and social communities to evaluate their relation with subjects' behavior.
We prove that social influence is correlated with these communities, and each one of them is (differently) significant for individuals. 
We define community-based features, which reflect the subject involvement in these groups, and we use them with a supervised learning algorithm to predict subject participation in social events. 
Results indicate that both communities and ego network are relevant sources of social influence, confirming that the ego network alone is not sufficient to explain this phenomenon.
Moreover, we classify users according to the degree of social influence they experienced with respect to their groups, recognizing classes of behavioral phenotypes.
To our knowledge, this is the first work that proves the existence of phenotypes related to the social influence phenomenon.
}

\section{Introduction}
\label{sec:intro}
The comprehension of human behavior is a complex and multifaceted problem. 
The understanding of the forces behind human beings actions can be beneficial to a growing number of applications, ranging from location-based services to mobile network management and epidemic spread prediction.
Many works in computational social science attempted to figure out the factors that affect people decisions and actions. We can classify them in three interlaced categories: physical location, homophily, and social ties.
\emph{Physical location} is a constraint in human movements, interactions, and relationships \cite{onnela2011geographic}. 
Previous works proved that the probability of being friends decreases with the geographic distance \cite{stewart1941inverse}.
\emph{Homophily} is the attitude of individuals to associate with subjects having similar interests, characteristics, and preferences \cite{mcpherson2001birds}. It can be viewed as a measure of similarity among subjects \cite{forster2012distance}.
The third influence factor is related to \emph{social ties}, which represent a key element in human behavior and groups formation. 

Communities and social groups are important structures in social networks.
The traditional approach in literature is to utilize the ego network to represent user's community in Online Social Networks (OSNs) and to analyze her/his behavior. 
An ego network is the social network of a subject, referred to as ``ego''. The other nodes, named ``alters'', are the subjects that have a direct social relationship with the ego, e.g., friends and relatives. 
Individuals not directly connected with the ego, e.g., friends of friends, are not considered in this network.
This approach does not take into account factors not embedded in network topology, which can originate communities, such as homophily, location, and social ties not included in the ego network.
However, all these factors contribute to people aggregation. 
Though geography poses limits in social interactions, physical locations act as aggregation points for activities, encounters, and meetings among people. 
Furthermore, subjects aggregate not only when they have friendship or kinship relation, but also when they have shared interests, according to the common identity and common bond theories \cite{back1951influence,ren2007applying}.
Some efforts have been made to analyze communities according to users' interests and visited locations. 
Nevertheless, most of the approaches consider each user as embedded in only one community.

Our hypothesis is that individuals are embedded in distinct communities related to the driving factors of human behavior described above.
We conceive each subject as part of communities that do not only consider her/his direct social relationships, but that involves complex aspects. Therefore, we analyze physical, homophily, and social communities. 
In particular, we focus on the social influence phenomenon and its correlation with human behavior. 
With social influence we indicate the attitude of a subject to be affected by other subjects' actions and decisions \cite{sunstein2002law}.
Social influence analysis can have a significant impact in prominent fields such as viral marketing and targeted advertisement.
This is confirmed by the intensive research that has been carried out during the last decade. 
Previous studies focused on two main directions. The former attempted to investigate and verify the existence of social influence \cite{anagnostopoulos2008influence,crandall2008feedback,la2010randomization,singla2008yes}. The latter aims to measure the influential strength, learning the probability of influence between pairs of friends \cite{goyal2010learning,tang2009social,saito2008prediction}.
In particular, a geo-social influence measure is introduced in \cite{zhang2012evaluating}, which combines geographic and social influence.
Differently from these works, our analysis focuses on the influence exerted by a group of people as a whole entity.
Therefore, we examine each community as a single source of social influence, instead of measuring influence probability among each pair of subjects. 
The rationale of this approach is based on the idea that subjects may follow the collective behavior of individuals $(i)$ living in their physical area, $(ii)$ with common interests, or $(iii)$ in their social community.

We make use of an Event-Based Social Network (EBSN) dataset to perform our analysis and to predict human behavior, in terms of user participation in social events. 
An EBSN is a web platform where users can create events, promote them, and invite friends to participate.
Events can range from small get-together activities, e.g., Sunday brunch or movie night, to bigger and complex events, e.g., concerts or conferences \cite{liu2012event}. 
Along with the event dissemination module, EBSN provides a social network service so as to connect friends and users with common interests.
The motivation behind the choice of utilizing an EBSN is related to the intrinsic agglomerative power of the events. In fact, participating in an event does not only represent a personal preference but also, a direct and explicit form of social interaction. 
Moreover, in the event main page a user can see the information related to the event, e.g., date, location, and description, along with the confirmed participants. This information may activate processes of social influence, which can drive user participation in the events \cite{georgiev2014call}. 
We summarize our contributions as follows:
\begin{itemize}
\item We introduce a novel interpretation of physical, homophily, and social community, as sources of social influence. 
Results indicate that both communities and ego network are relevant sources of social influence. Moreover, each user has a reference influence group, which is more correlated with her/his actions. 
\item We predict human behavior utilizing only the historical information related to the subject and to the members of her/his communities, achieving an average accuracy of 81\%. We prove that the ego network alone is not sufficient to achieve such performances. 
\item We classify users according to the degree of social influence they experienced with respect to their communities, discovering classes of correlations among them corresponding to a limited number of behavioral phenotypes \cite{poncela2016humans}.
\item Using the classification results, we predict subject's behavior considering only historical information related to the subjects belonging to her/his behavioral class. We achieve an average accuracy of 82\%, proving that user's historical data are not necessary if we know her/his behavioral class. This knowledge is also more privacy-preserving.

\end{itemize}

\section{Problem Definition}
In this work, we investigate the social influence phenomenon in order to characterize and predict human behavior.
We aim to examine the role of physical, homophily, and social communities as sources of social influence and evaluate their relation with subjects' behavior.
This purpose is accomplished by following two main directions.
We will refer to them as \emph{human behavior prediction} and \emph{user classification}.
The former targets to predict user participation in social events utilizing only historical information related to the user and to the groups she/he belongs to. 
The latter aims to distinguish users according to the degree of social influence they experienced with respect to their communities. 
Finally, we intend to utilize the results of user classification in the human behavior prediction.

\subsection{Dataset Description}
We make use of the EBSN \emph{Plancast} for our study. Plancast is a service for sharing upcoming plans with friends.
It allows users to subscribe each other providing direct connections among them. 
Subscription is similar to the concept of \emph{following} in OSNs, e.g., Twitter. The user can directly follow friends' event calendars: this mechanism allows the user to be aware of friends interests, events creation, and participation. 
When a user creates an event, subscribers are notified and can state their willingness to attend the event by RSVP (``yes'', ``no'' or ``maybe'').
Differently from Plancast, in EBSNs like \emph{Meetup}, users subscribe to groups according to their interests. Group membership does not imply any social tie between members, but only a common interest.
Thus, Plancast provides a more straightforward and clear way to recognize online social relationships compared to other EBSNs. 

We utilize a dataset \cite{liu2012event} that includes 93041 users and 401634 events, combined in 1702058 user subscriptions and 869200 user-event participations. 
The dataset also lists the location of the events, in terms of $(latitude, longitude)$ coordinates.
We restrict our analysis to the U.S., as most of the events have been organized there. We filter out users without any subscription and that attended less than 20 events. 
We set this threshold in order to build, per each user, a reasonable training and test set for the human behavior prediction, presented in Sect. \ref{results}. 
Moreover, we take into account only events with at least 20 participants.
We do not consider small events because we are interested in understanding users dynamics in large and complex contexts. Small events involve only cliques of friends and thus they have no relevance for this kind of analysis. 

\subsection{Community Extraction}
In this section, we clarify the elements involved in our study.
Let $E$ be the set of events and $V$ the users set, we define $\forall u \in V$:
\begin{itemize}
\item The set ${F}_u$, which includes the users subscribed by $u$.
\item The set ${A}_u$, which includes the events attended by $u$.
\item The \emph{centroid of interests} $c_u$, which represents the user's area of interests, based on the locations of the attended events. 
\end{itemize} 
Given that our dataset does not include any information related to the users' home location, we evaluate $c_u$ to assign a representative geo-location to the user.
This centroid is a pair of coordinates $(latitude, longitude)$, computed in two steps. 
First, we detect and remove the outliers from the coordinates of the events in ${A}_u$, according to the Median Absolute Deviation (MAD) measure.
Second, we average the coordinates of the remaining events in ${A}_u$ to compute $c_u$.

We now define three graphs that play a fundamental role in our analysis.
Let $SG$ be the undirected\footnote{We build an undirected graph according to similar works in literature.} $\mathbf{Social \ Graph}$, $SG=(V,E_s)$, where $E_s$ is the set of edges built utilizing the subscription connections in ${F}_u$.
We define the complete $\mathbf{Physical \ Graph}$ $PG=(V,E_p)$, where each pair of users $(i,j)$ is connected by an edge weighted by a function of the geographical distance between the users' $c_u$.
We utilize a Gaussian kernel as a function to transform the geo-distance into a measure of similarity.
We denote by $HG=(V,E_h)$ the complete $\mathbf{Homophily \ Graph}$, whose set of edges $E_h$ is weighted by an interest similarity measure defined as follows: $w^{HG}_{i,j}=(|{A}_i\cap {A}_j|/|{A}_i\cup {A}_j|)$, $ \forall i,j \in V$. This metric indicates the number of common events attended by each pair of users, normalized by the joint set of attended events. It represents how similar are two users in terms of interests. 

We aim to partition these three graphs in order to create a set of communities within each one of them: social communities for $SG$, physical communities for $PG$, and homophily communities for $HG$. The purpose is to group together users that have common social ties, which are in physical proximity, and which share interests, respectively. 
We partition users so as to maximize the \emph{modularity}.
As this is a NP-hard problem, we make use of a heuristic algorithm. We utilize the Louvain method \cite{blondel2008fast} for its computational efficiency.
Finally, we define the $\mathbf{ego \ network}$ $ego_u$, $\forall u \in V$. Every edge in the ego network represents a social tie between the ego and the alter. 
We extracted this network from $SG$. In fact, every node directly connected to $u$ in the $SG$ is an alter in $ego_u$.
Thus, $ego_u$ is a user-centered subgraph of $SG$.
Summarizing, we can see each user $u \in V$ as embedded in:
\begin{itemize}
\item Three different communities, each one representing a distinct behavior of the individual in three dimensions: social (ties), space (physical location), and interests (homophily). We will refer to these communities as $SC_u$, $PC_u$ and $HC_u$.
\item An ego network $ego_u$, representing her/his direct social ties.
\end{itemize}
A question can easily arise at this point of the study. Why do we consider both $ego_u$ and $SC_u$? The answer is straightforward by examining the nodes in each graph: the first includes only nodes directly connected to the user, the second includes additional nodes, according to the results of the Louvain partitioning method, and broader represents the social structure within the user is settled. 

\section{Empirical Analysis}
\subsection{Data preprocessing}
\label{preproc}
We define here the features that we utilize in both human behavior prediction and user classification.
From now on, we will use the term \emph{group} to indicate both the ego network and the three communities.
We build a dataset ${D}_u$ for each user $u \in V$, where each row $e$ in ${D}_u$ represents an event in ${A}_u$.
We call \emph{group participation} the feature $p^{g}_{e}(u)$ computed as follows:
\begin{equation}
\label{formula}
p^{g}_{e}(u)=\frac{|\{i \in g | e \in A_i \}|}{|g|}\ ,
\end{equation}
where $g \in \{ego_u,SC_u,PC_u,HC_u\}$ and $i \in V$. It indicates the number of users in group $g$ that attended event $e$, normalized by the dimension of the group. 
Thereby, each row in the dataset ${D}_u$ is a 4-tuple composed of the four features related to $p^{g}_{e}(u)$.
This feature represents a measure of group participation in an event. A high value indicates that most users in the group took part in the event.
The idea is that when a subject sees most of the group members confirmed their participation in an event she/he may be more willing to participate her/himself.

\subsection{Human Behavior Prediction}
\label{results}
In this section, we describe our method to predict human behavior, in terms of user participation in social events.
It should be noticed that our primary goal is not to find the best algorithm to perform prediction, but to show that the communities we identified above influence human behavior and can be used to predict subjects' decisions.
For this purpose, we utilize the dataset ${D}_u$ for each user $u \in V$. This dataset includes all the events attended by $u$. To make the dataset balanced, we append to the dataset $n$ rows corresponding to the events not attended by $u$, where $n=|{A}_u|$. We select the $n$ closest events to the centroid of interests $c_u$.
For each event, we compute the four features $p^{g}_{e}(u)$, as in (\ref{formula}).
Finally, the balanced dataset includes $2n$ events, equally distributed between attended and not attended events. Each row is composed of four features $\{p^{ego}_{e}(u),p^{SC}_{e}(u),p^{PC}_{e}(u),p^{HC}_{e}(u)\}$ and a target $T$, a Boolean variable indicating if the user attended the event or not.
The objective is to predict $T$. Therefore, we have to deal with a classification problem. 
From now on, we will indicate the features as $\{ego,SC,PC,HC\}$, to simplify the reading.

In order to limit overfitting and to reduce variability, we utilize a 10-fold cross-validation to split the dataset into training and test set. 
We build the folds so as to preserve the percentage of attended events in the dataset.
We employ three well-known machine learning algorithms to solve the classification problem: Decision Tree (DT), Support Vector Machine (SVM), and Deep Neural Network (DNN). 
We report in Table \ref{tab1} the performances related to these three algorithms, in terms of accuracy, precision, and recall.
SVM and DNN achieve the best performances.
We decide to continue our analysis utilizing SVM due to its computational efficiency.



\begin{table}
\caption{Prediction performances comparison}
\label{table1}
\centering
\begin{subtable}{.5\textwidth}
\centering
\begin{tabular}{M{1.5cm}M{1cm}M{1cm}M{1cm}}
\hline\noalign{\smallskip}
 & DT & SVM & DNN  \\
\noalign{\smallskip}\svhline\noalign{\smallskip}
Accuracy & 77\% & 81\% & 81\% \\    
Precision & 74\% & 84\%  & 85\% \\
Recall & 72\% & 76\% & 75\% \\
\end{tabular}
\subcaption{Classifier Performances}
\label{tab1}
\end{subtable}%
\begin{subtable}{.5\textwidth}
\centering
\begin{tabular}{M{1.5cm}M{1cm}M{1cm}M{1cm}M{1cm}}
\hline\noalign{\smallskip}
 & $ego$ & $SC$ & $PC$ & $HC$  \\
\noalign{\smallskip}\svhline\noalign{\smallskip}
Accuracy & 77\% & 77\% & 77\% & 76\% \\    
Precision & 81\% & 81\% & 82\%  & 82\% \\
Recall & 72\% & 71\% & 69\% & 67\% \\
\end{tabular}
\subcaption{Performances utilizing one fixed feature}
\label{tab2}
\end{subtable}
\end{table}

At this point, we investigate the relevance of each group in the subjects' behavior.
First, we try to evaluate our method utilizing only one feature among $\{ego,SC,PC,HC\}$.
The goal is to understand if one of them can be used for every user, without performance loss.
Results are shown in Table \ref{tab2}. It can be noticed that no feature achieves the previous performance, where all the features were considered. The ego network alone is not sufficient to obtain those results.
This proves that also $SC$, $PC$, and $HC$ are relevant in this analysis, confirming our hypothesis.

Second, we select one specific feature per user according to a feature selection algorithm.
We utilize the concept of mutual information to choose the feature. 
Table \ref{tab3} compares the average results of the three described scenarios: all the features (SVM results in Table \ref{tab1}), one fixed feature per every user (average results of Table \ref{tab2}), and one feature per user based on the feature selection algorithm.
\begin{table}
\centering
\caption{Prediction performances comparison: all features vs. one fixed feature vs. feature selection}
\label{tab3}       
%
%
\begin{tabular}{M{1.5cm}M{2.5cm}M{2.5cm}M{2.5cm}} 
\hline\noalign{\smallskip}
 & all features & one fixed feature & feature selection  \\
\noalign{\smallskip}\svhline\noalign{\smallskip}
Accuracy & 81\% & 77\% & 80\% \\    
Precision & 84\% & 82\%  & 84\% \\
Recall & 76\% & 70\% & 76\% \\
\noalign{\smallskip}\hline\noalign{\smallskip}
\end{tabular}
\end{table}

Feature selection closely approaches the performance obtained utilizing all the features.
Certain users are mainly influenced by their physical community, some others by their social community, some by their homophily community, and others by their direct social relationships, i.e., their ego network.
We also notice that the selected features are equally distributed among the users, i.e., each feature is selected for almost 25\% of the users. 
This evidence confirms the importance of all the four groups, and it further validates our hypothesis.
In fact, each community is differently significant for the users and each of them has a reference influence group, which is more correlated with her/his actions.

\subsection{Communities Analysis and User Classification}
\label{classification}
We examine here the subjects' behavior in terms of social influence with respect to the groups she/he belongs to. 
For this purpose, we utilize ${D}_u$ (as described in Sect. \ref{preproc}), and we compute for each user $u \in V$ the average of the features $p^{g}_{e}(u)$ over all the attended events.
We call these values \emph{group influence} $i^{g}_{u}$.
Each user is identified by four features, each one of them representing the degree of influence the group has in the user participation history.

\begin{figure*}
        \centering
        \begin{subfigure}[b]{0.32\textwidth}
               \includegraphics[width=\textwidth]{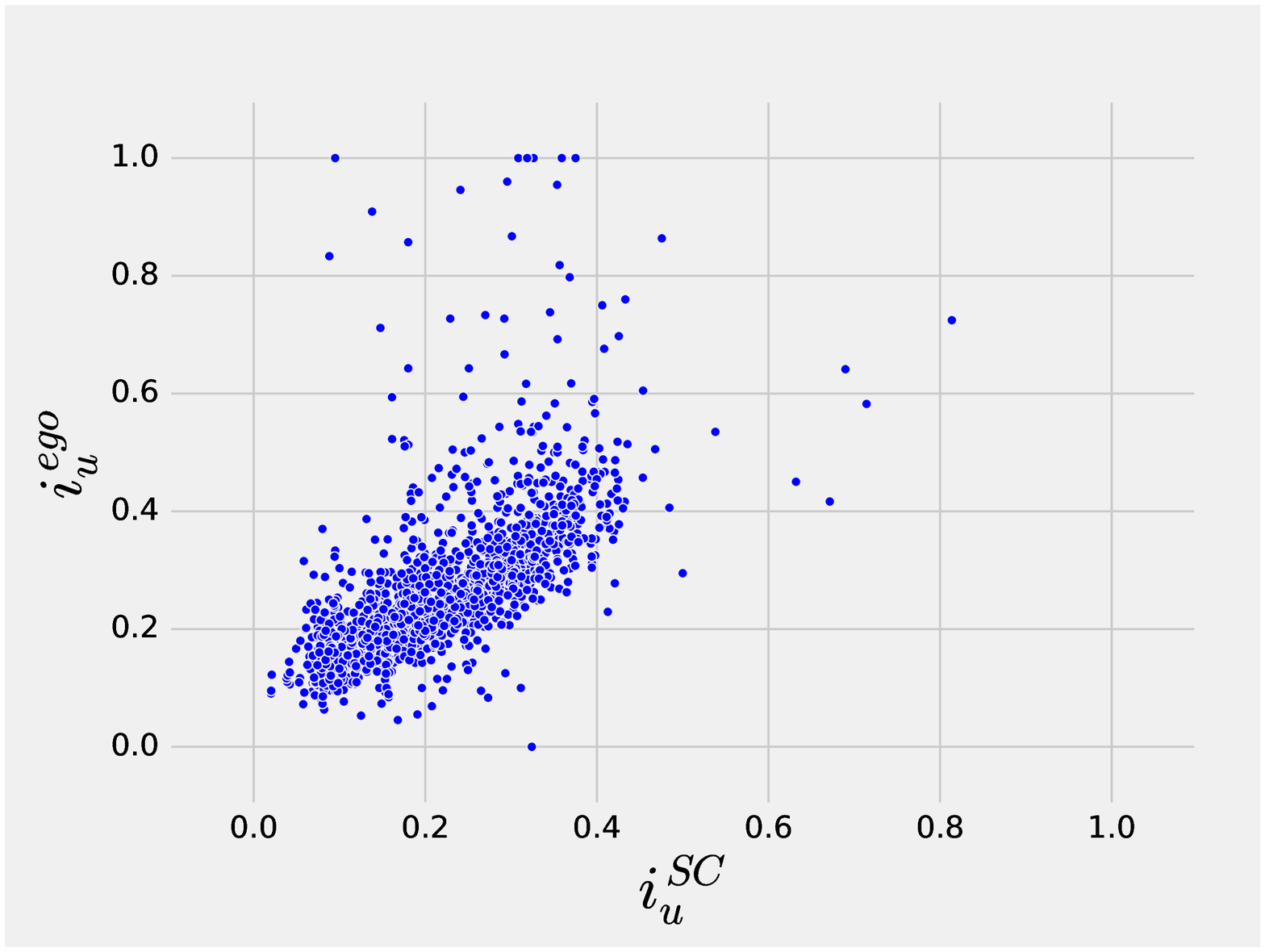}
                \caption{$i^{ego}_{u}$ vs. $i^{SC}_{u}$}
                \label{fig1a}
        \end{subfigure}\hfill%
        \begin{subfigure}[b]{0.32\textwidth}
               \includegraphics[width=\textwidth]{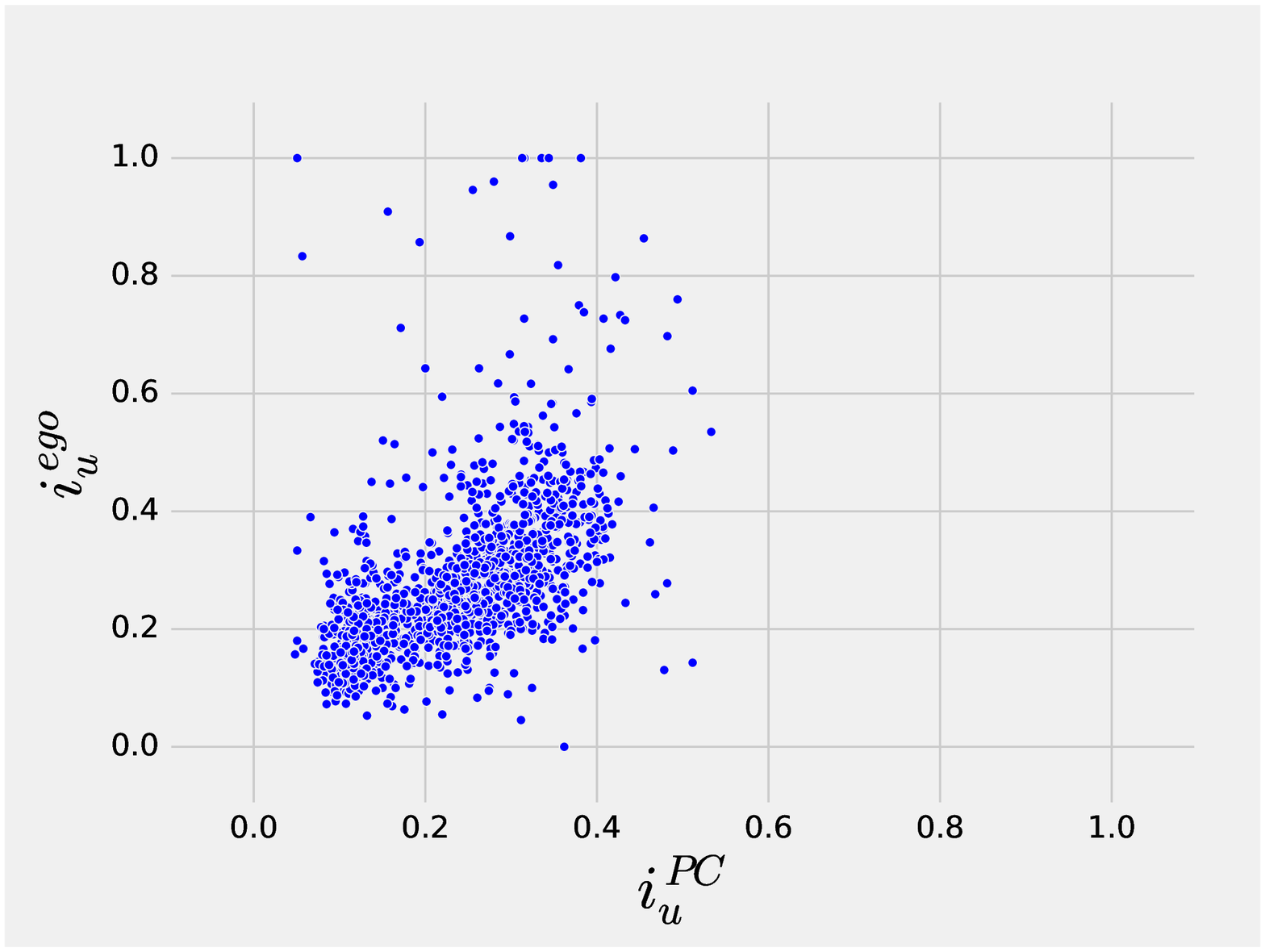}
                \caption{$i^{ego}_{u}$ vs. $i^{PC}_{u}$}
                \label{fig1b}
        \end{subfigure}\hfill%
        \begin{subfigure}[b]{0.32\textwidth}
                \includegraphics[width=\textwidth]{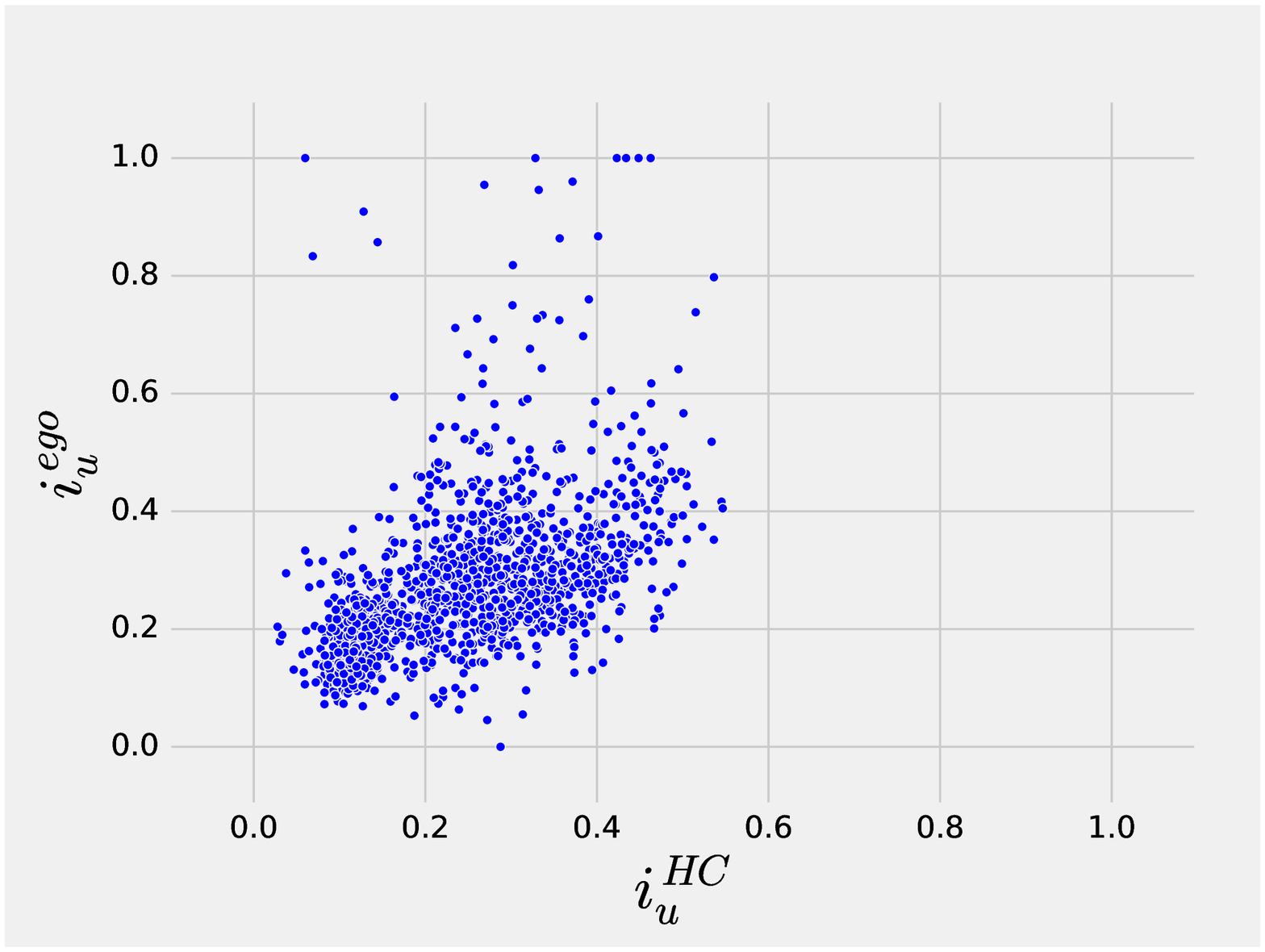}
                \caption{$i^{ego}_{u}$ vs. $i^{HC}_{u}$}
                \label{fig1c}
        \end{subfigure} \\
        \vspace{0.06in}
         \begin{subfigure}[b]{0.32\textwidth}
               \includegraphics[width=\textwidth]{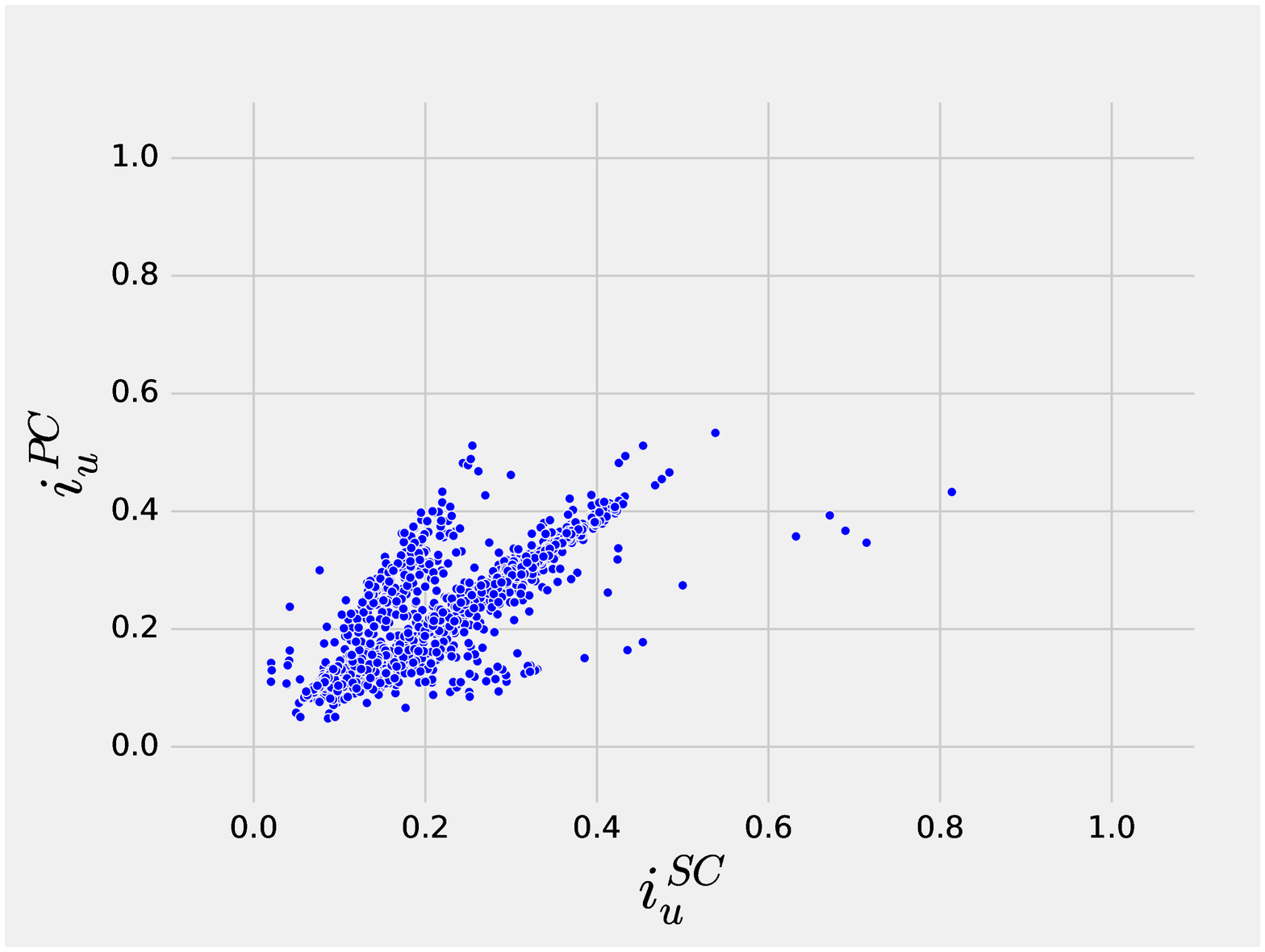}
                \caption{$i^{PC}_{u}$ vs. $i^{SC}_{u}$}
                \label{fig1d}
        \end{subfigure}\hfill%
        \begin{subfigure}[b]{0.32\textwidth}
               \includegraphics[width=\textwidth]{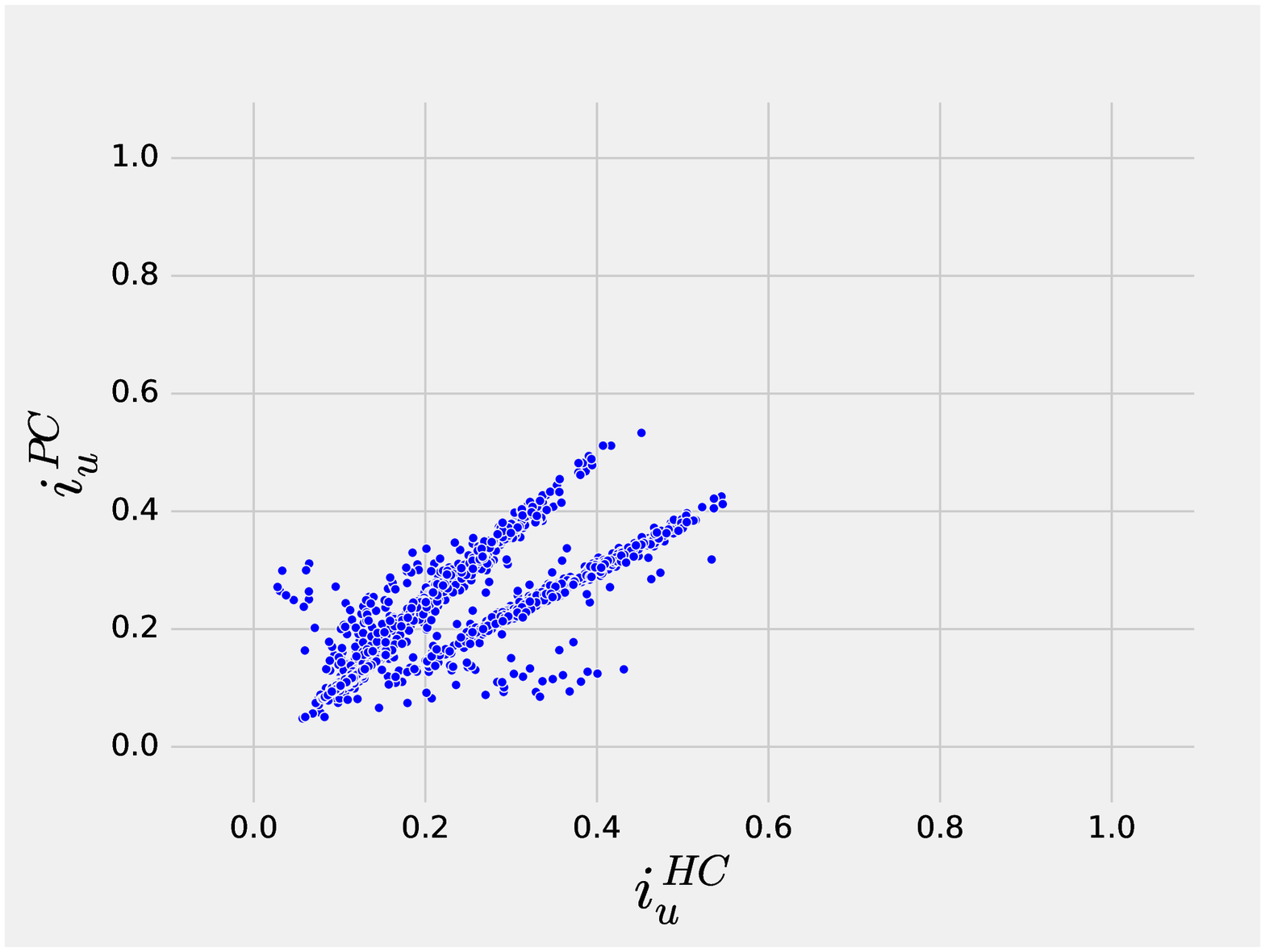}
                \caption{$i^{PC}_{u}$ vs. $i^{HC}_{u}$}
                \label{fig1e}
        \end{subfigure}\hfill%
        \begin{subfigure}[b]{0.32\textwidth}
        
                \includegraphics[width=\textwidth]{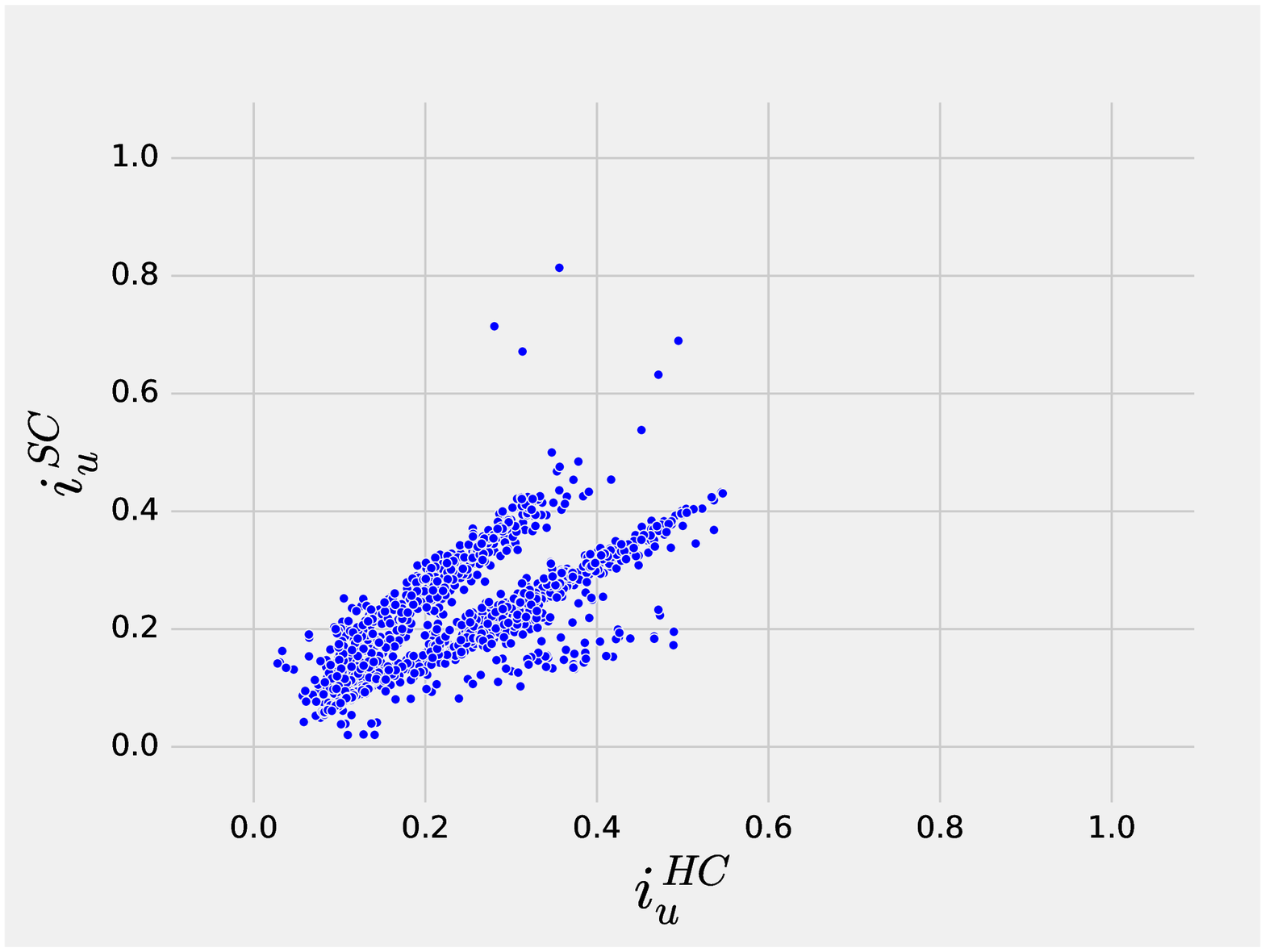}
                \caption{$i^{SC}_{u}$ vs. $i^{HC}_{u}$}
                \label{fig1f}
        \end{subfigure}        
        \caption{Combination of values related to $i^{ego}_{u}$, $i^{SC}_{u}$, $i^{PC}_{u}$, and $i^{HC}_{u}$. Each point represents a user.}
        \label{fig1}
\end{figure*}

Our aim now is to understand if there exists any relation between these features.
We start analyzing the ego network. 
Figures \ref{fig1a}, \ref{fig1b}, and \ref{fig1c} relate $i^{ego}_{u}$ with the values of the other three groups.
Each point in the figures indicates a user. The y-axis represents the values related to the ego network parameter, while the x-axis changes with the group under inspection. 
We observe scattered correlation patterns between the ego network feature and the other three features $(i^{SC}_{u}, i^{PC}_{u}, i^{HC}_{u})$. 
As expected, the strongest correlation can be observed between $i^{ego}_{u}$ and $i^{SC}_{u}$, as the ego network is a subgraph of the $SG$. 
We then evaluate the combination of the three features related to the communities, as shown in Figs. \ref{fig1d}, \ref{fig1e}, and \ref{fig1f}. Every picture presents similar patterns between each other, but different distributions and stronger correlations with respect to the figures above.
It can also be appreciated the presence of structures that indicate classes of correlation between every pair of features.

\begin{figure*}
        \centering
        \begin{subfigure}[b]{0.45\textwidth}
               \includegraphics[width=\textwidth]{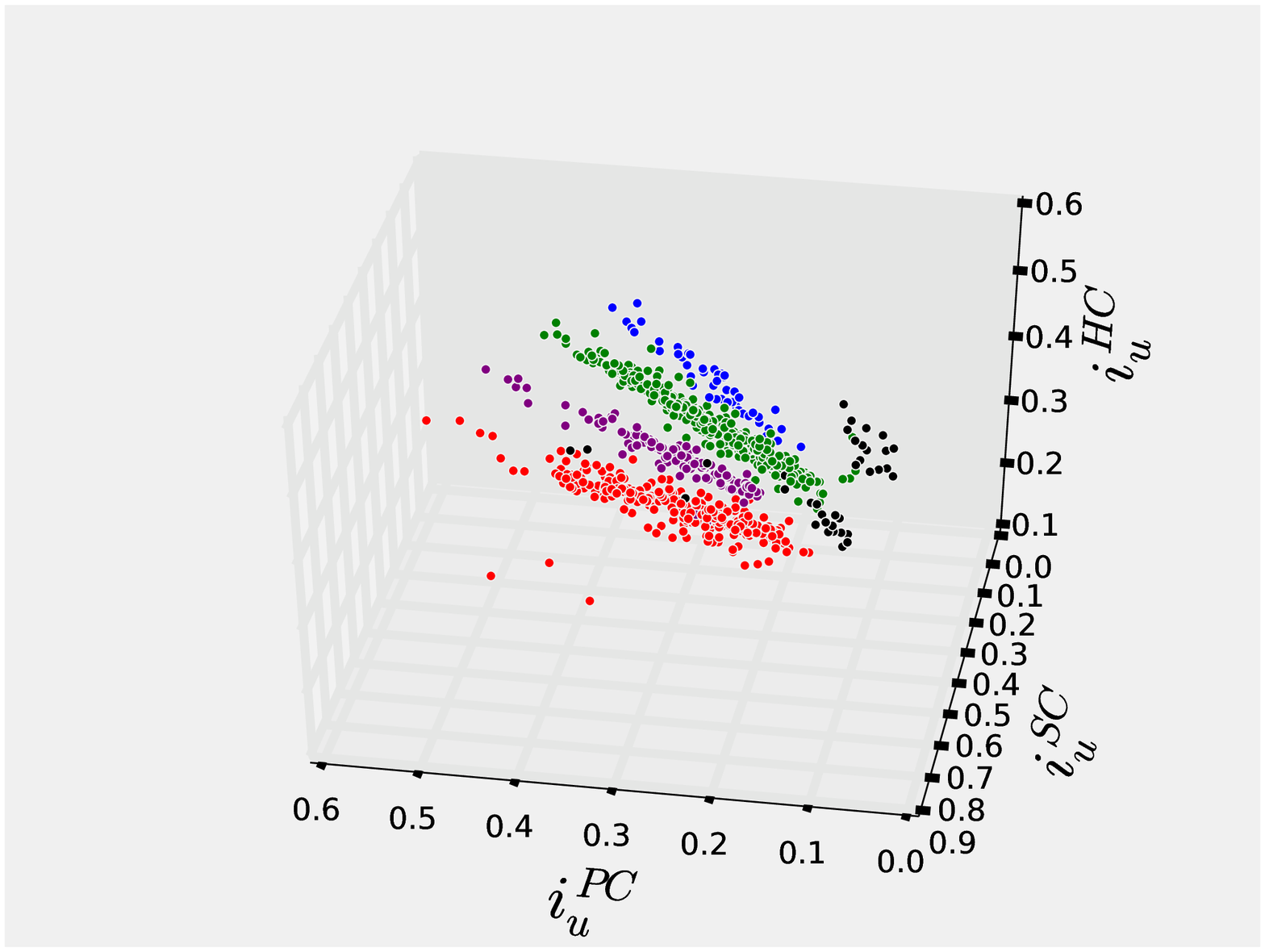}
                \caption{Fingers clustering} 
                \label{fig3}
        \end{subfigure}\hfill%
        \begin{subfigure}[b]{0.45\textwidth}
               \includegraphics[width=\textwidth]{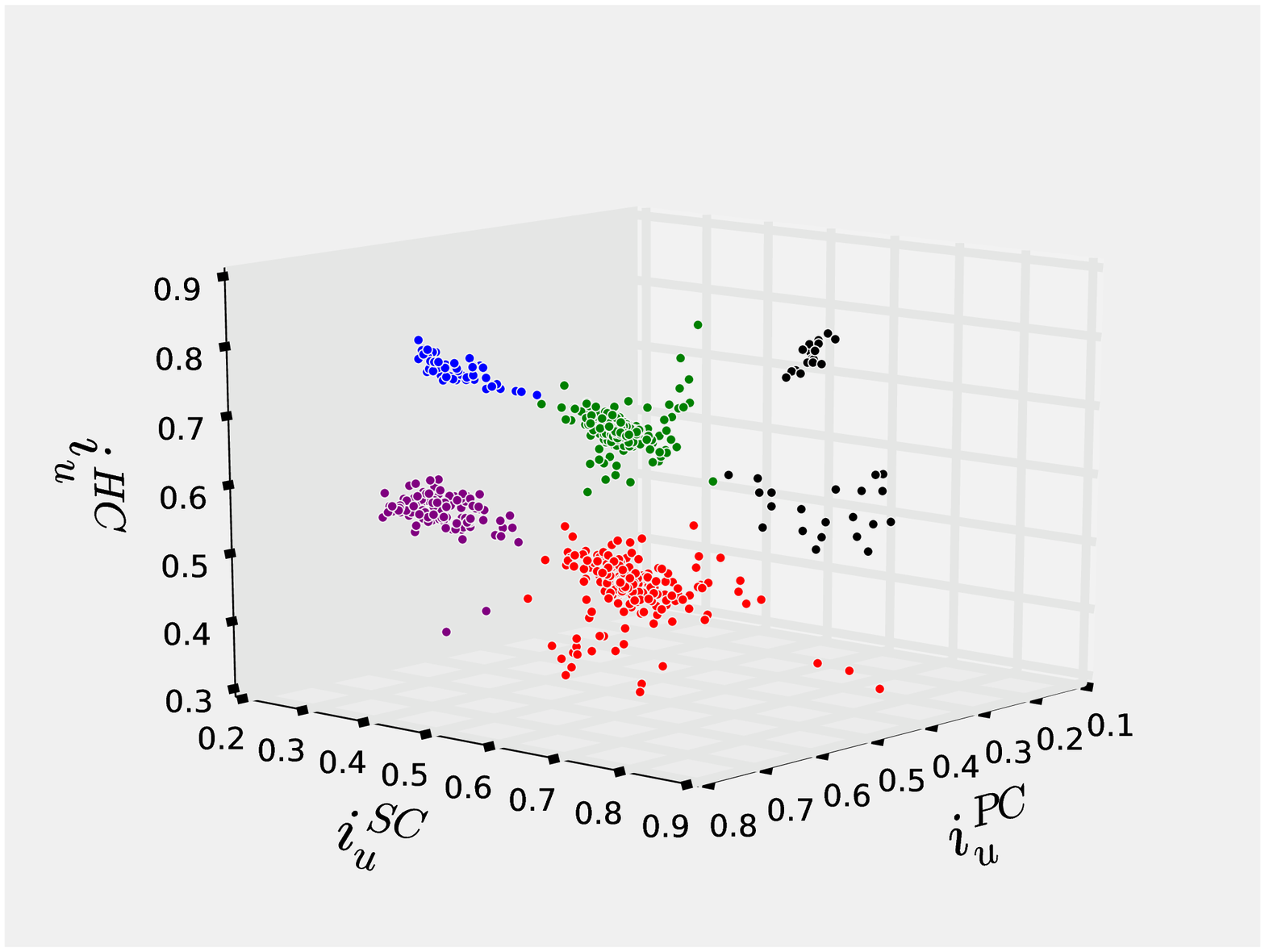}
                \caption{Fingers clustering on the unit sphere} 
                \label{fig4}
        \end{subfigure} \\ 
        \vspace{0.06in}
        
        \begin{minipage}{0.45\textwidth}
        \caption{Fingers and influence classes: each point represents a user in the 3-D space ($i^{SC}_{u}$,$i^{PC}_{u}$,$i^{HC}_{u}$). In Figs. \textbf{(a)} and \textbf{(b)}, each color distinguishes a finger. In Fig. \textbf{(c)}, each color distinguishes a class of influence.}
        \end{minipage}
    \hfill
     \begin{subfigure}{0.45\textwidth}
     \includegraphics[width=\textwidth]{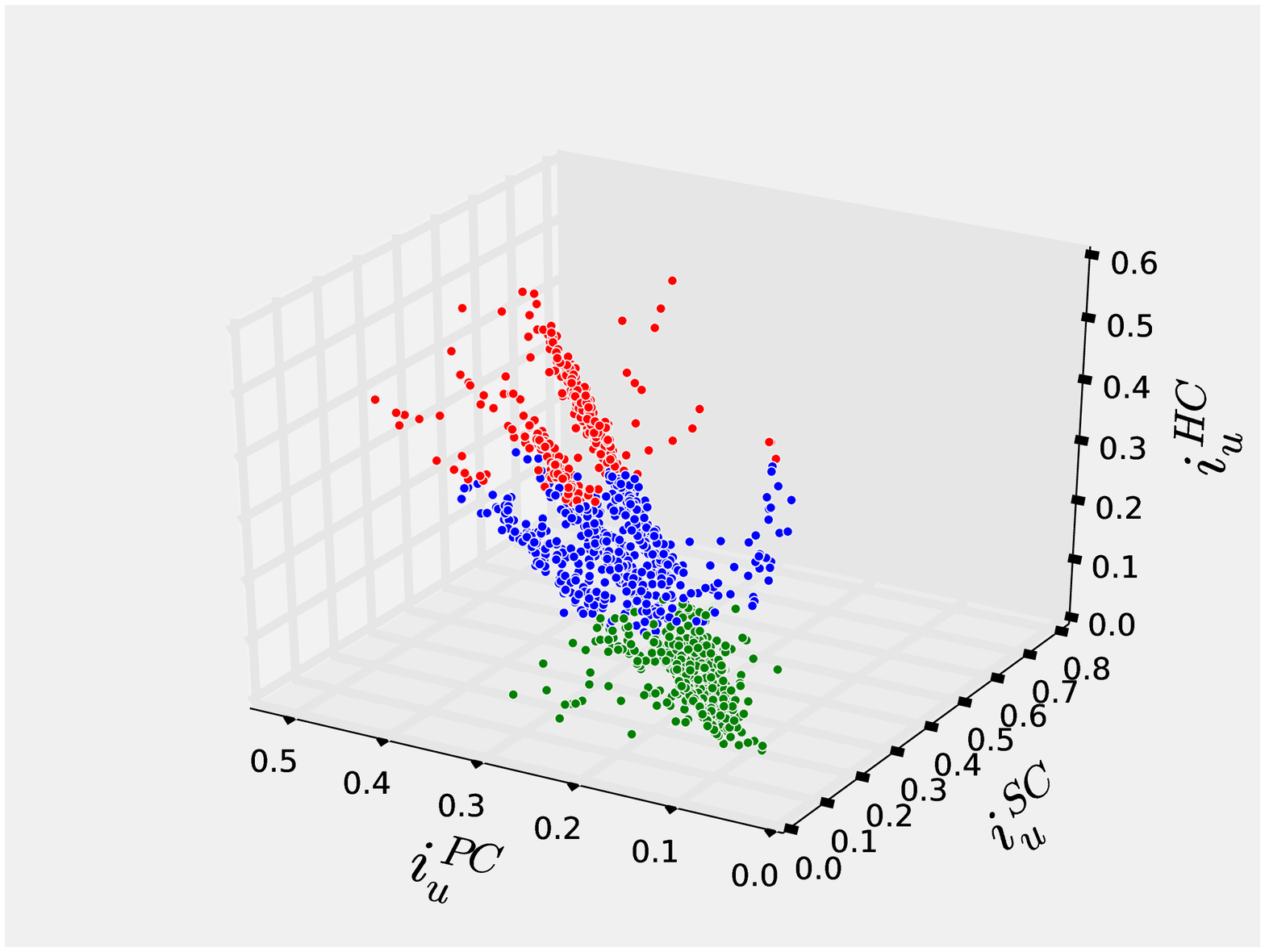}
	\caption{Classes of influence clustering}
	\label{fig5}
    \end{subfigure}
\label{figall}   
 \end{figure*}

To further investigate and to better understand these structures, we observe and evaluate the three features in a three-dimensional space, as depicted in Fig. \ref{fig3}. 
For visualization purposes, we do not plot users with values close to zero, otherwise this 3-D figure would have been unreadable in two-dimensions. 
We can clearly detect structures that look like five fingers.
These fingers show a three-fold correlation between the features. 
This correlation further confirms the interdependence among the three driving factors of human behavior previously described.
Every user in a finger has the same type of behavior in terms of percentage of influence among the communities. In fact, the ratios between feature values are constant over the finger.
This unexpected clustered pattern shows ratios between features that do not vary in $\mathbb{R}^3$, but that assume discrete values in the three-dimensional space. 
It should be noticed that these fingers overlap with each other when projected onto a 2-D plane. This is the reason why we recognize only two fingers per plot in Figs. \ref{fig1d}, \ref{fig1e}, \ref{fig1f}. 

We observe that each finger can reasonably be conceived as a line passing through the origin. 
To cluster together points on the same finger, we convert the Cartesian coordinates to spherical ones $(r,\theta,\phi)$ and
we project each point on the surface of a unit sphere. 
This procedure reveals five separate clusters, as depicted in Fig. \ref{fig4}.
In such a way, we need only the angles $(\theta,\phi)$ to distinguish the points on the five fingers.
We utilize the k-means algorithm to perform the partitioning. The number of clusters $k$ has been chosen according to the number of visible fingers. Clustering results are shown in Figs. \ref{fig3} and \ref{fig4}. 
The colors in the plots distinguish each finger and reveal five classes of users, which from now on we will call \emph{fingers}.


Furthermore, we propose to classify users based on whether they are subject to ``low'', ``medium'', or ``high'' degree of social influence.
We refer to these partitions as \emph{influence classes}.
We use the k-means algorithm ($k=3$) to cluster the users into these classes. 
Results are depicted in Fig. \ref{fig5}, which includes also users with feature values close to zero (differently from Fig. \ref{fig3}).
Red points represent users highly influenced by the groups, while green points indicate the low influence class. 

The two partitions presented above reflect two different properties in the spherical coordinates $(r,\theta,\phi)$.
Each user is characterized by a specific combination of the three features, with the radial variable $r$ characterizing the overall level of social influence, and the angular variables $(\theta,\phi)$ describing specific ratios between pairs of community-features. 
The level of social influence is determined by many factors related to the users, e.g., the involvement in the communities, the personal preferences, the social relationships, and the geographic location. 
As a consequence, we observe a distribution of values between a minimum and a maximum with a broad dispersion around the mean value. We would have expected a similar distribution also for the angular variables. 
Instead, we observe a multimodal distribution composed of a small number of centroids, i.e., the five fingers, with a narrow distribution around each of them. 
These statistical patterns are probably a clue of some interesting sociological and psychological factors driving human behavior.
This result confirms that subjects' behavior can be described by a limited number of behavioral phenotypes \cite{poncela2016humans}. To our knowledge, this is the first work that proves the existence of behavioral phenotypes related to the social influence phenomenon.

Finally, we try to utilize the above classification results in the human behavior prediction.
More specifically, 
we want to exploit user's class for predicting event participation, training a model only with historical data related to subjects belonging to the same class of the user. With the term class we refer both to fingers and influence classes.
We treat each class separately, utilizing a subset of the users in the learning process and the remaining users in the testing phase. We utilize a 10-fold cross-validation also for this purpose. 
Table \ref{tab4} reports the prediction results related to both the influence classes (low, medium, high), and fingers ($f_1,f_2,f_3,f_4,f_5$). 
As a baseline, to be compared with these results, we need to evaluate the prediction performance considering all the users as belonging to the same class. Thus, we repeat the same methodology considering a unique class of users. In this instance, we achieve an accuracy of 72\%, a precision of 75\%, and a recall of 66\%.
\begin{table}
\caption{Performances based on classes of influence (\textbf{a}) and fingers (\textbf{b})}
\label{tab4}
\centering
\begin{subtable}{.5\textwidth}
\centering
\begin{tabular}{M{1.2cm}M{1cm}M{0.6cm}M{1cm}M{0.6cm}}
\hline\noalign{\smallskip}
 & \textbf{average} & low & medium & high  \\
\noalign{\smallskip}\svhline\noalign{\smallskip}
Accuracy & 80\% & 72\% & 82\% & 87\% \\    
Precision & 80\% & 71\% & 82\%  & 88\% \\
Recall & 71\% & 57\% & 73\% & 82\% \\
\end{tabular}
\subcaption{Classes of influence}
\label{sub-table-1}
\end{subtable}%
\begin{subtable}{.5\textwidth}
\centering
\begin{tabular}{M{1.2cm}M{1cm}M{0.65cm}M{0.65cm}M{0.65cm}M{0.65cm}M{0.65cm}}
\hline\noalign{\smallskip}
 & \textbf{average} & \textbf{$f_1$} & \textbf{$f_2$} & \textbf{$f_3$} & \textbf{$f_4$} & \textbf{$f_5$}  \\
\noalign{\smallskip}\svhline\noalign{\smallskip}
Accuracy & 78\% & 80\% & 78\% & 78\% & 77\% & 79\% \\    
Precision & 79\% & 80\% & 75\% & 83\%  & 74\% & 82\% \\
Recall & 67\% & 67\% & 69\% & 68\% & 65\% & 66\% \\
\end{tabular}
\subcaption{Fingers}
\label{sub-table-2}
\end{subtable}
\end{table}

We can observe that:
\begin{itemize}
\item The highly influenced class achieves the best performance. This is reasonable and it was also predictable because every feature is a measure of social influence. Thus, the features fit perfectly in case of high influence. For the same reason, the low influence class performs poorly compared to the other two classes.
\item Fingers do not exhibit the same behavior of influence classes. There is no finger that outperforms the others. This is also expected because every finger represents a class of correlation, and as such, it includes also users belonging to the low influence class.
\item The baseline results are close to the ones related to the fingers. It seems that finger prediction gets perturbed from users belonging to different influence classes.
\end{itemize}

These considerations gave us the idea to combine classes and fingers in $3\times5$ sub-classes, named as behavioral classes. In such a way we expect to take benefits from both partitions. The results, reported in Table \ref{tab6}, confirm our idea.
\begin{table}
\centering
\caption{Performance of the prediction based on behavioral classes}
\label{tab6}       
%
%
\begin{tabular}{M{1.5cm}M{1cm}M{1cm}M{1cm}M{1cm}}
\hline\noalign{\smallskip}
 & \textbf{average} & low & medium & high  \\
\noalign{\smallskip}\svhline\noalign{\smallskip}
Accuracy & 82\% & 74\% & 83\% & 89\% \\    
Precision & 84\% & 75\% & 86\%  & 92\% \\
Recall & 79\% & 73\% & 78\% & 85\% \\
\noalign{\smallskip}\hline\noalign{\smallskip}
\end{tabular}
\end{table}

The behavioral classes outperforms all the previous results. The improvement is also reflected in the influence classes: all of them gain in performance with respect to the outcomes in Table \ref{sub-table-1}.
This improvement is due to the further partitioning introduced by the fingers. In such a way, similar users are grouped together according to the two properties described above.
We demonstrate how event participation history is not necessary if we know user's behavioral class. This knowledge has also more predictive power compared to user history (Table \ref{tab3}) and it is also more privacy-preserving.
In fact, in this scenario, we do not utilize any information related to the user during the learning phase, but only historical data related to the subjects in the same behavioral class. 

\section{Conclusions}
In this work, we investigated human behavior and in particular the relation between social influence and communities.
We proved that the ego network alone is not sufficient to explain this phenomenon.
Other groups, such as physical, homophily, and social communities, are also relevant sources of social influence. 
Results show that each user is mainly influenced by one of these groups. 
We classify users according to the degree of social influence they experienced with respect to their groups, recognizing a limited number of behavioral phenotypes.
These phenotypes are probably a clue of some interesting sociological and psychological factors, which are going to be further investigated with similar datasets in our future works.
\bibliographystyle{spmpsci.bst}
\bibliography{bibl}

\end{document}